%% file: qandies_plain-v9.2.2.tex
\documentclass[runningheads, envcountsame, a4paper]{article}
\usepackage{amssymb}
\usepackage{mathtools}
\usepackage{dsfont}
\usepackage{mathrsfs}
\usepackage[pdftex]{graphicx} 
\usepackage[hyphens]{url}
\usepackage{epstopdf} 
\usepackage{microtype} 
\usepackage{dcolumn}
\usepackage{verbatim,comment} 
\usepackage{float}
\usepackage{color}
\usepackage[utf8]{inputenc}
\usepackage{appendix}
\usepackage{slashbox}
\usepackage{multirow,hhline}
\usepackage[table]{xcolor}
\usepackage{makecell}
\usepackage{authblk}
\usepackage{braket}

\providecommand{\keywords}[1]
{
	\small	
	\textbf{\textit{Keywords---}} #1
}

\begin{document}

\title{Quantum information and beyond --- with quantum candies}

\author[1]{Junan Lin}
\author[2]{Tal Mor}
\author[2]{Roman Shapira}

\affil[1]{Institute for Quantum Computing and Department of Physics and Astronomy, University of Waterloo, Waterloo, Ontario N2L 3G1, Canada}
\affil[2]{Computer Science Department, Technion –- Israel Institute of Technology, Technion city, Haifa 3200003, Israel}

\date{\today}

\maketitle
\setcounter{footnote}{0} 

\input{abstract}
\input{content}

%
%
\bibliographystyle{plain}
\bibliography{candyQKD}

\end{document}

%% file: abstract.tex
\begin{abstract}
	The field of quantum information is becoming more known to the general public.
	However, effectively demonstrating the concepts underneath 
	quantum science and technology to the general public can be a challenging job.
	We investigate, extend, and greatly expand here ``quantum candies'' 
	(invented by Jacobs), a pedagogical model for intuitively describing some 
	basic concepts in quantum information, including quantum bits, 
	complementarity, the no-cloning principle, and entanglement.
	Following Jacob's quantum candies description of the well-known quantum key
	distribution protocol BB84, we explicitly demonstrate additional quantum cryptography protocols and
	quantum communication protocols, 
	using generalized quantum candies (including correlated pairs of qandies).
	These demonstrations are done 
	in an approachable manner, that can be explained to
	high-school students, without using the hard-to-grasp 
	concept of superpositions and its mathematics.
	
	The intuitive model we investigate has a fascinating overlap with 
	some of the most basic features of quantum theory. Hence, it can be a 
	valuable tool 
	for science and engineering educators who would like to help 
	the general public to gain more insights into quantum science and technology. 
	
	For the experts, the model we present, due to not employing
	quantum superpositions,
	enables --- in some sense --- 
	extending far beyond quantum theory. 
	Most remarkably, ``quantum'' candies
	of some unique type can be defined, such that non-local boxes (of the
	Popescu-Rohrlich type) as well as regular (correlated) quantum candies can be
	generated by a single ``quantum'' candies machine.
	
\end{abstract}
\keywords{Quantum Information, Quantum Cryptography, Quantum Candy, Physics Education, Nonlocal Boxes}

%% file: content.tex
\section{Introduction}

Quantum information science and technology is a growing field that provides interesting concepts such as quantum computers, quantum teleportation, and quantum cryptography.
Its development has proved fruitful in both theoretical and experimental aspects, leading to various demonstrations in research labs. More recently the hi-tech became majorly involved, with companies such as IBM, Google, Intel, Microsoft, and Alibaba (plus many startup companies) investing in quantum devices for computing and/or communication and cryptography. 

The relevant ``quantum concepts'', in their simplest forms, can be demonstrated with simple information units known as quantum binary digits (a binary digit gets one of two values, zero or one), named quantum bits or qubits.
Many applications of quantum cryptography, including quantum key distribution (QKD) protocols~\cite{bennett1984quantum,bennett1992quantum,Ekert1991quantum}, can be explained using qubits.
While the basic ideas behind quantum information in general, and especially behind these QKD protocols, are not very complicated for an ``insider'', they can be abstruse for someone without the necessary physics or mathematics background.
This hinders general audiences from correctly understanding these concepts.
An intuitive description of such concepts will potentially be beneficial not only for the general public but also for researchers in other fields to understand and visualize protocols. 

\subsection{A Brief History of the Model}

In this work, we follow --- and greatly expand --- a simple but powerful model of ``quantum candies'' (or in short, ``qandies'') originally suggested by Jacobs~\cite{Kayla2009}, who proposed a new type of very special candies along with unusual machines that produce those special candies.
Jacobs' qandies are generalized here so that correlated pairs of qandies can be generated and measured.
The non-conventional properties of the quantum candies and those machines --- plus our wide and wild expansions --- are used here as explanatory tools for various quantum concepts and technologies. 

The model had originally been developed in two (independent) steps: Karl Svozil presented ``chocolate balls'' in several papers (e.g.~\cite{svozil2006staging}), including using these on an actual stage to demonstrate a pseudo-quantum model that resembles quantum cryptography. 
Kayla Jacobs, independently~\cite{Kayla2009}, invented an intuitive and much-closer-to-quantum variant based on properties of hypothetical candies having two different colors and two different tastes. 
To the best of our knowledge, Jacobs' model was presented in seminars (at MIT and at the Technion, and mainly to high-school students) but has never been officially published.

The qandies model does not require employing the concept and mathematics of
superpositions. We name it ``qandies'' but a more formal name for this model may
be ``superpositionless quantum theory''. Such a theory has a lot in common
with conventional quantum theory, yet it is far from being identical to
quantum theory, as we we shall see especially in
Sections~\ref{Sec:nontrivial-exten} and~\ref{Sec:NLB}.  

A preliminary version of this paper (by J.L.~and T.M.) appeared in \cite{lm20}.

\subsection{The Structure of the Paper}

The paper is structured as follows.
In Section~\ref{Sec:candy-section} we review Svozil's and Jacobs' demonstrations of the most well-known QKD protocol: the BB84 protocol~\cite{bennett1984quantum}. 
We then propose several straightforward extensions of the basic qandy model in Section~\ref{sec:trivial-exten}. 
Non-trivial (i.e., more sophisticated) 
extensions are suggested in Section~\ref{Sec:nontrivial-exten} --- extensions that 
compare and/or distinguish the qandies model from standard quantum theory.
Such extensions are of more interest to the experts and less so to the more general
public.

In Section~\ref{Sec:qandies-concepts} we expand the quantum candies model to
explain various quantum concepts including entanglement, and an attack on
``quantum bit commitment'' in Subsection~\ref{sec:bit-commitment}.
In Section~\ref{Sec:new-meas} we introduce new measurement capabilities via ``Pseudo-Bell" measurements and provide qandy protocols using these concepts: BHM96~\cite{bhm96} and MDI-QKD.
In Section~\ref{Sec:qandy-gates} we define single-qandy and multi-qandy gates,
and give several usage examples, including the qandy version of the famous
``Superdense coding" protocol in Subsection~\ref{sec:superdense}.
In Section~\ref{Sec:SQKD} we discuss in detail a specific extension 
of the qandy model that models semi-quantum key distribution protocols.
In Section~\ref{Sec:NLB} we present the most non-trivial 
extension of the qandies model --- an extension that, 
surprisingly, contains Non-Local Boxes and
quantum-theory-related candies as two special cases, hence majorly  
distinguishes the qandies model from standard quantum theory.
We end with some brief closing remarks in Section~\ref{Sec:Discussion}.

\section{Quantum Cryptography, Chocolate Balls, Quantum Candies, the BB84
Protocol, and Beyond}\label{Sec:candy-section}
Suppose that two users, Alice and Bob, would like to communicate secretly.
They may do so if they share a secret random and sufficiently long key (typically, a string of bits).
Thus, the remaining problem is how to distribute an identical random and secure key.
Quantum key distribution (QKD) protocols solve the key distribution problem by utilizing quantum objects that, due to the rules of quantum theory, cannot be copied, and any attempt to copy them or even slightly obtain information from them enables Alice and Bob to detect the presence of the eavesdropper.
Thus an eavesdropper can block the communication but cannot learn the secrets. 

The BB84 QKD protocol proposed by Bennett and Brassard in 1984~\cite{bennett1984quantum}, is simple and it can also serve as an excellent tool for understanding the basics of quantum theory.
Svozil and Jacobs found simple \emph{and sweet} ways to describe the BB84 protocol in somewhat classical ways.

\subsection{Chocolate Balls and Generalized Urn Model}\label{GU section}

Svozil~\cite{svozil2006staging} described a simple way to stage the BB84 protocol using real-life chocolate balls, based on the so-called generalized urn (GU) model by Wright~\cite{wright1990generalized}.
In this model, the qubits are represented by chocolate balls wrapped in black foil with two binary numbers printed on the surface, one binary digit (0 or 1) in red color and the other binary digit (0 or 1) in green.
If we denote the case of a zero written in red and a zero written in green as \{0,0\}[red,green], then the four types of balls also include \{0,1\}[red,green], \{1,0\}[red,green] and \{1,1\}[red,green].

Any user can view the numbers on the chocolate balls, but \emph{must obey} the restriction that he/she must put on red-filtering or green-filtering glasses first so that only one binary number can be seen every time a person looks at a ball. 

The chocolate balls are drawn from a large urn containing an equal number of the four types of balls.
To send a string of random bits to Bob, Alice wears a pair of colored glasses, randomly draws a single ball from the urn, and records the number she sees so she keeps the data regarding both the digit and its color. 
The ball is then sent to Bob who, before obtaining it, randomly picks a glasses color and wears the glasses. 
Only then Bob is allowed to observe it, and then to record the color and number (as Alice did).

After repeating these steps as many times as they wish, each time again randomly picking a new pair of colored glasses, Alice and Bob then announce their choices of colors publicly and keep the bits obtained only in the cases where they picked the same color of glasses, which guarantees that they will obtain the same bit string. 
About half the data is thrown away, in case all previous choices were made at random.

If the transmission between Alice and Bob for the chocolate balls is not being eavesdropped, or if eavesdropped by an eavesdropper that \emph{follows the rules}, then the distributed key can be trusted. 
However, a cheating eavesdropper (we call her Eve) can obtain \emph{all} the transmitted information by simply not obeying the imposed law of wearing colored glasses.
Moreover, her observations do not change the chocolate balls.
Svozil~\cite{svozil2014non} analyzed various urn models and their connections to quantum physics, and of course was aware that this demonstration of BB84 only partially illustrates the BB84 protocol, as it does not illustrate its true quantumness or the resulting security.




\subsection{Jacobs' Quantum Candies Model}\label{BB84 section}
We now consider Jacobs' model, where, instead of chocolate balls, there exist candies to which we gradually assign ``quantumness''.
We call the resulting sweets ``quantum candy'' (or, \emph{qandy}, or Jacob's qandies).
The model was originally presented by Kayla Jacobs~\cite{Kayla2009} as an easy way to present quantum bits and QKD to the general audience.

Jacobs considered candies that have two different \emph{general properties}, color and taste.
The color of a candy can be red (R) or green (G), while the taste can be chocolate (C) or vanilla (V). 

At first, the above general properties of color and taste and specific properties (red, green, chocolate, vanilla) show much similarity to Svozil's sweets, and may actually even found to be fully equivalent! Here are the four options of chocolate balls versus four options of candies, with an obvious one-to-one correspondence:
\begin{gather*}
  \{0,0\}\text{[red,green]} \longleftrightarrow \{C,R\}\text{[taste,color]}\\
  \{0,1\}\text{[red,green]} \longleftrightarrow \{C,G\}\text{[taste,color]}\\
  \{1,0\}\text{[red,green]} \longleftrightarrow \{V,R\}\text{[taste,color]}\\
  \{1,1\}\text{[red,green]} \longleftrightarrow \{V,G\}\text{[taste,color]}
\end{gather*}
It is trivial to see that this translation from the generalized urn model is exact thus far, hence if an eavesdropper (on BB84) could \emph{both} look and taste, the candies protocol becomes as insecure as the generalized urn model once Eve does not obey the rules. 

\begin{figure}[!ht]
	\centering
	\includegraphics[width=0.55\textwidth]{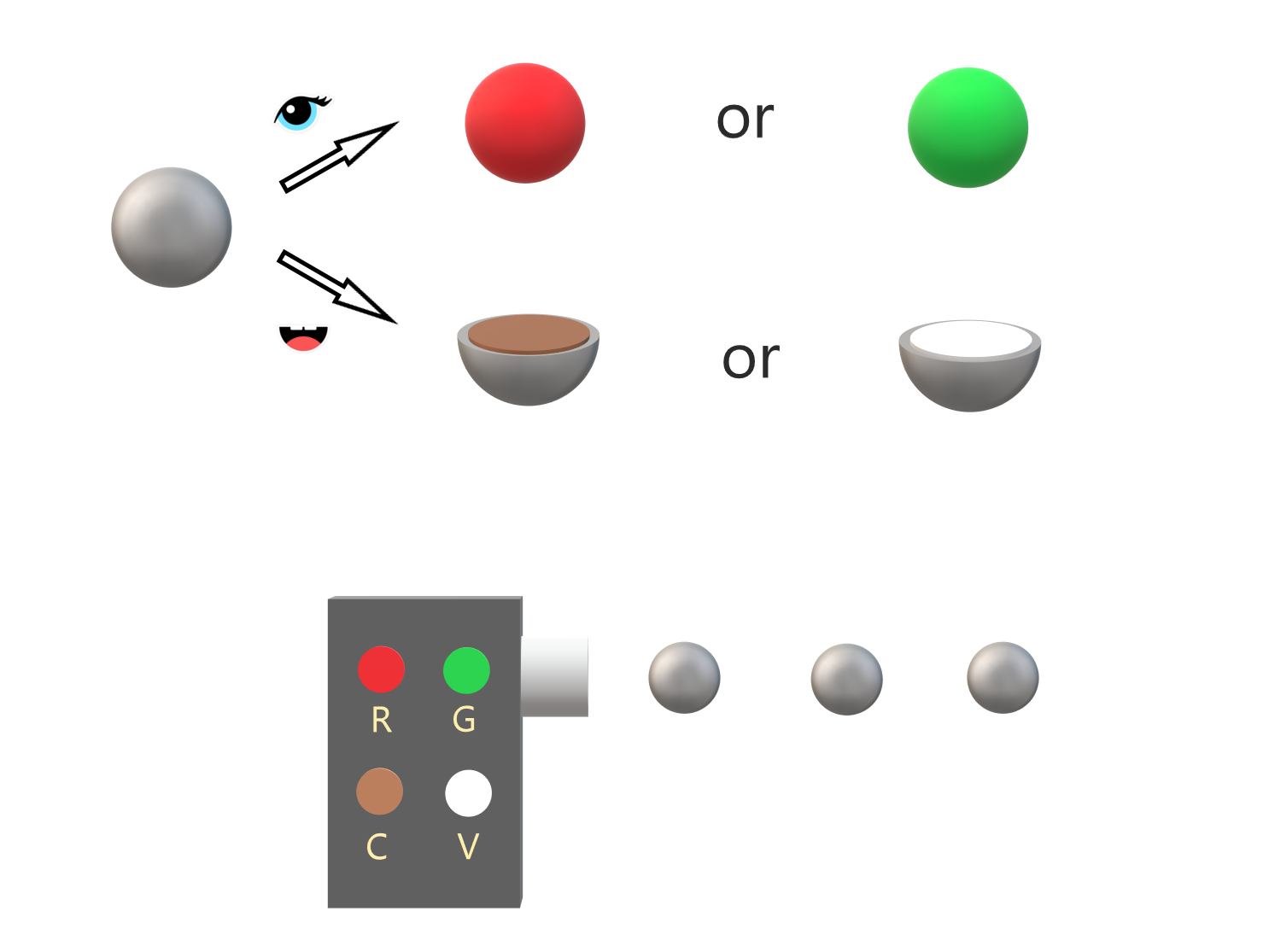}
	\caption{
	Top: illustrative figures of qandies with two general properties -- color and taste.
	Before any choice of observation, all qandies appear identical. If one looks at a qandy, it will appear as red or green, but one cannot taste it again; if one tastes a qandy, he/she would taste either chocolate or vanilla, but one cannot then learn anything about its color.
	Bottom: a qandy-producing machine.
	The user can press one of the four buttons on the machine and generate the corresponding qandy. Only one general property can be fixed for every qandy produced, while the other general property would be randomly assigned by the machine (if we follow Einstein) or have the other property non-existing at all (if we follow modern understanding and Jacob's qandies).
	} 
	\label{candy-fig}
\end{figure}

As described in Figure~\ref{candy-fig}, we impose some unusual rules onto these candies that make them ``\emph{qandies}''.
First, each qandy has only a single specific property:
\begin{equation*}
  \{C\}\text{[taste]},\ \{V\}\text{[taste]},\ \{R\}\text{[color]},\ \{G\}\text{[color]}
\end{equation*}
What are the implications of this rule in terms of generating the qandies and in terms of ``observing'' them?
First, a user can \emph{only} learn about one general property of a qandy, but not both. Namely, if one looks at a qandy, it will appear as red or green, but one cannot taste it anymore (its taste is destroyed by looking at it); if one tastes a qandy, he/she would taste either chocolate or vanilla, but one cannot learn anything about its color anymore (its color is destroyed).
Second, a qandy-making machine would have four buttons, one for each specific property $\{R, G, C, V\}$ (a user generating the qandy cannot choose more than one specific property).
Third, if the machine generates (as an example) a chocolate qandy ($\{C\}$), one can taste and know it is chocolate, but if one looks, a random color is seen ($\{R\}$ or $\{G\}$) and the taste is destroyed.

The key feature of these qandies is that each single qandy \emph{really} has only a single specific property.
This is one form of what is known as the complementarity principle\footnote{Other forms of the principle exist, e.g., that elementary particles like electrons or photons have both particle characteristics and wave characteristics.} in quantum physics: if color is defined, taste cannot be defined, and if taste is defined color cannot be defined. The rule of complementarity applies both to the person (i.e. person/machine) preparing the qandies, and to anyone observing the qandy.

But maybe an extremely sophisticated (having some future technology) eavesdropper can somehow know both the color and taste?
Surprisingly, Einstein's view of quantum physics, e.g. his famous sentence ``God does not play dice'' and also the EPR paradox~\cite{EPR35}, is also in complete correspondence to the above.
In his view quantum physics is incomplete: it has ``hidden variables'' and a deeper theory (not yet known to science, but maybe known to that futuristic Eve) may unveil the hidden variables and hence will provide both the color and the taste of these candies.  

If we try to use our ``classical imagination'' to \emph{build} the machine, we might think as Einstein did: imagine a futuristic machine that produces standard classical candies that are wrapped with an identical piece of edible but opaque candy paper, such that none of the candy's properties can be known by only looking at a freshly prepared candy.
The machine is a black box to all users, who can only interact with the machine through four external buttons: $\{C\},\{V\},\{R\},\{G\}$.
If Alice presses $\{R\}$ or $\{G\}$ to prepare a candy with definite color, that candy does not have a defined taste, and \emph{vice versa} --- when one general property is well defined, the other general property becomes \emph{random}. 
So any user cannot fix or learn both the color and taste of a candy.
And indeed, if Alice produces a candy with a specific color, say $\{R\}$, and Bob tastes that candy, he will taste a random taste ($\{C\}$ or $\{V\}$). 
And similarly, looking at a candy having a specific taste will result in seeing a random color ($\{R\}$ or $\{G\}$).

Einstein believed that even if one general property must be randomized, it's actually not --- that specific property is still \emph{there}, but well-hidden. 
Hence a \emph{deeper theory} (beyond quantum) could reveal both general properties with their explicit specific properties, so the candies eventually will follow (in the eyes of a futuristic scientist) classical intuition and eventually --- will be identical to Svozil's chocolate balls.

Today --- due to a theorem known as Bell's theorem~\cite{bell1964einstein}, and various confirming experiments (the first done by Aspect~\cite{Aspect1982,Aspect1982a}) --- the modern view is that we do not follow Einstein's belief anymore. 
As understood today, when Alice asks for a qandy with a definite taste, the machine would produce one with Alice's desired taste, but with \emph{no defined color}, and if Alice wants a qandy with a definite color, the machine would produce one with Alice's desired color, but with \emph{no defined taste}! 
The complementarity rule is so deeply inherent in Jacobs' qandies model that we may say the other general property does not exist at all (it is not even random), and it only becomes random if an observation is made.


Finally, the qandy model illustrates that an observation alters the properties of the qandies, leading to a no-cloning principle and to secure QKD.

\subsection{Orthogonality and No-Cloning}\label{Subsec:no-cloning}
We now discuss an important implication of these qandies, which will be useful in explaining the BB84 protocol. 
Here, the key concept is \emph{orthogonality}: we say that two qandy states are orthogonal to each other when they are different states of the same general property:
$\{R\}$ and $\{G\}$ are orthogonal states, and $\{C\}$ and $\{V\}$ are also orthogonal states.
Crucially, orthogonal states can be \emph{perfectly distinguished} if an observation is made to distinguish them. 
I.e., if Alice sends to Bob a color qandy, and tells Bob it is a color qandy, then Bob can distinguish whether the qandy is red or green by looking at it. 
On the other hand, if Bob is given randomly a red or chocolate qandy, then no observation can determine, with 100\% chance of being correct, the specific property of that qandy.

The ability to identify a particular state directly translates to the ability of copying (cloning) a qandy.
Thus, the existence of non-orthogonal states results in the impossibility to copy a random qandy that can be in arbitrary states.
This is named the no-cloning principle and is central to the safety of many quantum cryptography protocols, as we will now see.

\subsection{BB84 QKD with Qandies}\label{BB84-qandy}
We now demonstrate how Alice and Bob may utilize qandies to achieve the task of secure key distribution, using the BB84 protocol.
First, prior to sending anything, they determine a set of rules that assign binary values to states of each specific property; for example, they may assign 0 to red color and chocolate taste, and 1 to green color and vanilla taste.
This would allow their communication to result in a string of bits.
Then, when they are later separated and would like to exchange a key to encrypt their message, they repeatedly perform the following: Alice randomly presses one of the four buttons on her qandy machine, records her choice, and has the qandy delivered to Bob.
Upon receiving, Bob randomly decides what he would like to do with it: he can either ``look'' and record its color, or ``taste'' and record its taste.
He then translates the result into 0 or 1 and records the bit value \emph{as well as} his action.

After this has been repeated a sufficient number of times, Alice and Bob will communicate through the classical channel to compare their preparation and observation \emph{methods} for every qandy\footnote{The classical channel is assumed to be insecure yet unjammable.}.
Specifically, Alice tells Bob whether she prepared a qandy with definite color or a definite taste, and Bob tells Alice whether he tasted or looked.
Note that they do not reveal their bit values to each other.
They keep the result if the two methods match, and discard otherwise\footnote{If Bob can keep the qandies un-measured in a qandies memory, the bit rate can be doubled, as he will taste or look only after learning Alice's preparation method.}.
For example, if for qandy number 17 Alice generated a qandy with definite color (say $\{R\}$), and Bob measured a general property of a color, they will share an identical bit (unless natural noise or an eavesdropper destroyed it and introduced an error).
This establishes a string of random bits between the two parties. 

The safety of this protocol can be understood from the no-cloning principle.
Let us analyze a specific eavesdropping attack --- the ``measure-resend attack'':
Suppose Eve gambles and looks at qandy number 17. She will see $\{R\}$ and hence get full information. 
She then creates a new copy of Alice's $\{R\}$ qandy and sends it to Bob, and Bob observes $\{R\}$ as he is supposed to.
However, we know that, in general, learning or cloning is impossible as discussed earlier; indeed, if Eve gambles badly and decides to taste qandy number 17, she will learn irrelevant information (a random taste) and resend a definite taste, hence ``corrupt'' the original bit that Alice intended to send. With a probability of 50\% Bob will see $\{G\}$ instead of $\{R\}$, i.e., Eve generated noise.

From Table~\ref{BB84-table} we see that overall, Eve's measure-resend attack will result in about $25\%$ mismatched bits among those shared between Alice and Bob, which shall be revealed to them by comparing a sufficiently long subset of their bits (called TEST bits) through the classical (insecure yet unjammable) channel.
In this case, Alice and Bob can discard their shared bits, and re-start this procedure again (via a different qandies channel) to obtain a new bit string, and verify (as we saw) that no eavesdropper is present.
It has been proven that a secret key can be obtained via post-processing of the data, as long as the error-rate is smaller than around 11\%~\cite{Shor2000, abb+14, sbc+09}. 

Lastly, we note that even if the error-rate is smaller than the threshold, Alice and Bob's bits are still not entirely identical and not entirely secret (Eve may have partial information on the bits).
In order to resolve these issues, Alice and Bob must take two additional steps called \emph{Error-Correction} and \emph{Privacy Amplification}, described in \cite{bbbss92, bbb+06, abb+14, sbc+09}.

\begin{table}[!ht]
	\centering
	\caption{
	Agreement probability between Alice and Bob due to the measure-resend attack.
	The table entries indicate actions taken by the eavesdropper as well as their occurring probabilities.
	A green cell implies agreement between Alice and Bob, whereas a red cell implies disagreement.
	In 25\% of the cases, their bits do not agree.
	}	
	\begin{tabular}{ |c|c|c|c|c| } 
		\hline
		\backslashbox{Bob}{Alice} & \multicolumn{2}{|c|}{Color} & \multicolumn{2}{|c|}{Taste} \\ 
		\hhline{*{5}{-}}
		\multirow{2}{5em}{Look} & \multicolumn{2}{|c|}{ \cellcolor[HTML]{9AFF99} Look, 50\% } & \multicolumn{2}{|c|}{\multirow{2}{*}{Discard}} \\
		\cline{2-3}
		\multicolumn{1}{|c|}{} & \cellcolor[HTML]{FFCCC9} Taste, 25\% & \cellcolor[HTML]{9AFF99} Taste, 25\% & \multicolumn{2}{|c|}{} \\
		\hline
		\multirow{2}{5em}{Taste} & \multicolumn{2}{|c|}{\multirow{2}{*}{Discard}} & \cellcolor[HTML]{FFCCC9} Look, 25\% & \cellcolor[HTML]{9AFF99} Look, 25\% \\ 
		\cline{4-5}
		& \multicolumn{2}{|c|}{} & \multicolumn{2}{|c|}{ \cellcolor[HTML]{9AFF99} Taste, 50\%} \\
		\hline
	\end{tabular}

	\label{BB84-table}
\end{table}

\subsection{A Few Direct Extensions}\label{sec:trivial-exten}

Two simple variants of BB84 are presented to show that making the properties somewhat more classical is possible, while keeping the goal of secure communication.
If qandies had only the color property and not the taste (or vice versa --- if had only the taste), a classical description would be immediate, cloning would be possible, and all parties, the sender the receiver, and any potential eavesdropper will see everything --- the green or the red. 
One variant of BB84, which is somewhat close to being classical, works when taste exists but qandies with taste are only rarely used. 
Such a protocol can still be proven secure~\cite{Lo2005}.
Another variant suggested the possibility that the tastes exist, but the machine used by the sender can only generate two colors (red and green) and just one taste (say, chocolate). 
As long as the receiver can detect both colors and both tastes, this protocol is secure~\cite{Mor1998}. 


A much more well-known protocol, the B92 protocol (when adjusted to qandies), uses just a single color and a single taste. 
In the B92 protocol~\cite{bennett1992quantum}, Alice only prepares two types of qandies: red and chocolate, which she and Bob agreed to stand for bits 0 and 1 respectively.
When Bob receives a qandy, he randomly chooses to taste or look, but does not reveal his choice to Alice as in BB84.
Instead, he publicly announces to Alice whether he \emph{succeeds} or \emph{fails} for every qandy, where the success condition is defined as \emph{tasting vanilla} or \emph{seeing green}.
He then associates seeing green with the bit 1 because Alice never prepared green and red could not lead to seeing green (only chocolate could), and he associates vanilla with 0 because Alice never prepared vanilla and chocolate could not lead to tasting vanilla (only red could).
After sending many qandies, as in BB84, they compare a subset of the bits to determine whether an eavesdropper is present, and final post-processing is used to generate a final key.

Lastly, we can consider various QKD protocols in which Alice has the full capabilities of generating and measuring qandies previously described, whereas Bob is ``color-blind" -- namely he cannot look at qandies.
The motivation for such a constraint is semi-quantum key distribution (SQKD) protocols \cite{bkm07, bgkm09}, where one party is quantum and the other party is classical.
Under the ``Color-blind Bob" constraint, the minimal set of operations available to Bob include:
\begin{itemize}
	\item
		Taste a qandy
	\item
		Prepare a qandy with a specific taste 
	\item
		Reflect a qandy obtained from Alice back to her, undisturbed
\end{itemize}
We discuss this topic in greater detail in Section~\ref{Sec:SQKD}.

\section{Non-Trivial Extensions of Qandies}\label{Sec:nontrivial-exten}

Some extensions beyond two colors and two tastes may lead to
non-triviality: options that exist for qandies, yet not for quantum entities, or
vice versa.
Namely, some characteristics of qandies as well as some 
extensions of qandies are not necessarily consistent with quantum physics.
This section focuses on such extensions; their trivial and non-trivial
options. Non-experts might prefer to skip this section. Section~\ref{Sec:NLB}
goes even much further in terms of far-from-quantum-physics extenstions.



\subsection{Mixed States of Qandies}\label{Subsec:mixed}
\begin{figure}[!ht]
	\centering
	\includegraphics[width=0.55\textwidth]{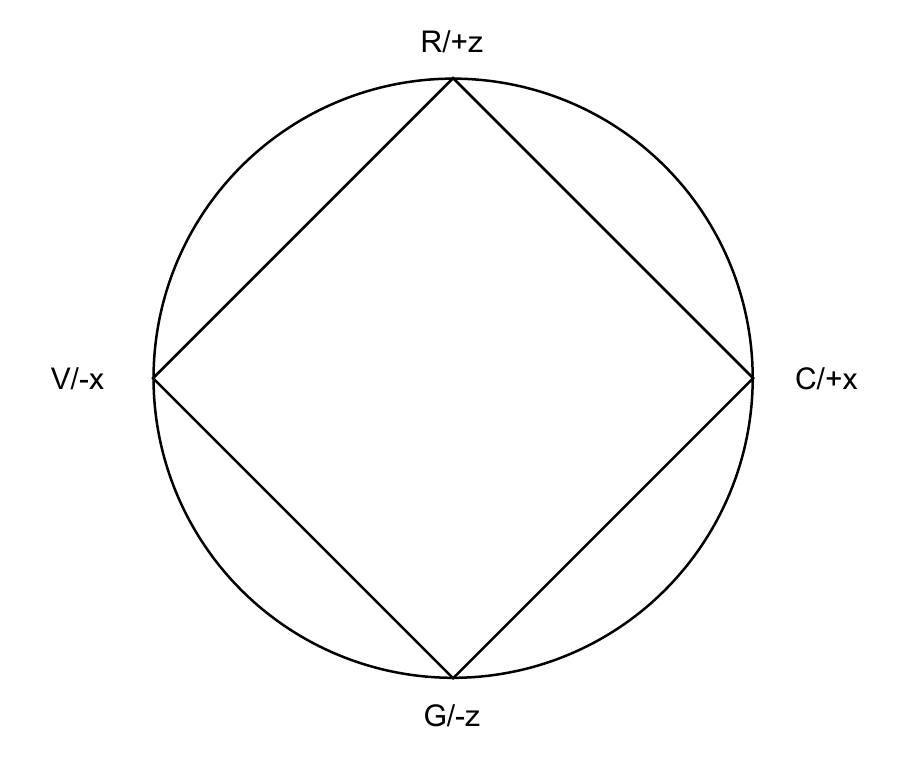}
	\caption{
		A qubit ``Bloch circle'' vs. a qandy ``Bloch square".
	} 
	\label{fig:qandy-bloch}
\end{figure}

We may visualize a single qandy on a circle:
the two colors 
correspond to the $\pm z$ points and the two tastes correspond to the $\pm x$
points (WLG). 
Any mixture of a color qandy and a taste qandy can be placed on one of the sides of the square (see Figure~\ref{fig:qandy-bloch}).
For example, a mixture of $\{V\}$ with probability $p$ or $\{G\}$ with probability $1-p$ is represented by a corresponding point on the line connecting $\{V\}$ and $\{G\}$.

We conjecture that points \emph{inside} the square also represent mixing of qandies, e.g. the middle point of the square may be represented (in the simplest case) as an equal mixture of colors or as an equal mixture of tastes.
This is similar to the notion of a density matrix of a single qubit in quantum theory, where \emph{one} density matrix (i.e. one point inside the circle) can represent various probability distributions of pure quantum states.
We leave the proof or refutation of this conjecture to future work.

Lastly, we note that in quantum theory, in contrast to the qandy case, the entire area of the circle is meaningful (in addition to the square area).
We leave extensions of qandies that enable more than four points on that circle and enable approaching quantum theory (once the proper limit is described) to a future paper.

\subsection{Beyond Two Colors and Two Tastes}
We can easily extend the qandies model to have more than two colors 
and tastes leading to additional key distribution protocols; 
some of those are actually ``beyond quantum'' in one sense or another.

For example, if instead of trying to imitate quantum bits--qubits, we try to imitate quantum trits--qutrits, we simply add one additional color (say, blue) and one additional taste (say peanut butter).
Interestingly, if we follow restrictions imposed by quantum theory, we must have the same number of colors and tastes.
Of course, Alice and Bob are not forced to use all three colors and tastes in such a case, but once we define the candies to be analogous to qutrits, the eavesdropper can make use of that extended qandies space. 

When using qandies, nothing prevents us from defining (as an example)
three colors and two tastes, or vice versa\footnote{A comment for the experts:
Of course, this kind of model is not consistent with the notion of a Hilbert space.}.

An even more fascinating extension is to go back to just two colors and two
tastes, but add a third property, say texture, to yield ``the six-state
protocol'', and (if considering mixed states as well) to extend the
abovementioned square inside a circle (see Section~\ref{Subsec:mixed}) 
to a whole cube inside a whole
sphere (known as the Bloch sphere).

The six-state protocol is a generalization of the BB84 protocol, 
where 3 pairs of orthogonal states (each pair belonging to one general property) 
of qandies are used; 
this requires a slightly more general machine that is capable of preparing 
qandies with 3 general properties, a color, a taste, and a texture --- 
the texture of a qandy can be, for example, soft or crunchy.
As in BB84, the machine can only prepare a qandy with one definite 
specific property, and one can only learn about one general property of a qandy. 
And if a qandy is prepared with one specific property (say the taste of vanilla) an attempt to look at it or to ``feel'' its texture must yield a random outcome, and the person tasting it feels no softness or crunchiness.
The resulting QKD protocol is (in some sense) more secure than BB84:
if we assume Eve applies an observation and sends according to her outcome, then among the bits intercepted by Eve, for which Alice and Bob used the same general property, $1/3$ of them would result in an inconsistency between Alice and Bob.
This is higher than in the BB84 protocol, meaning that there exists a level of natural error such that BB84 cannot be made secure while the six-state protocol can be made secure. 
Of course, this comes at some cost --- increasing the total number of qandies that Alice must share with Bob, due to the increased number (i.e. $2/3$) of discarded cases.

This extension is fascinating --- 
interestingly, having just two specific properties (two colors, etc.) 
and having four general properties --- e.g. color, taste, texture, and smell, 
is not consistent with the rules of quantum 
theory\footnote{A comment for the experts:
This topic, named mutually unbiased bases (MUBs), 
is also non-trivial in quantum theory.
For instance, if we define $n$ colors and $n$ tastes, for $n$ equal
2, 3, 4, or 5, it has been proven 
that we can define at most $n+1$ general properties, yet for $n=6$
it is not clear if one can define as many as 7 general properties.}.
We can, however, define such a qandy model consistent with Quaternionic Computing.
Quaternionic computing (as described in \cite{fs03}) is an alternative version of quantum computing, which uses ``quaterbits'' (instead of ``qubits'') - two-level systems with quaternionic (instead of complex) amplitudes.
Quaternions are an extension of complex numbers, where instead of a single imaginary unit $i$ there are three such units - $i, j, k$ - and the following relations hold:
\begin{equation*}
	i^2 = j^2 = k^2 = ijk = -1.
\end{equation*}
Using the definition, one can show that a single ``quaterbit'' of the form $\ket{\psi} = \alpha\ket{0}+\beta\ket{1}$, with $\alpha, \beta$ quaternions of the form $q = a+ib+jc+kd$, has 5 mutually unbiased bases (MUBs).
This corresponds to a single qandy with 5 possible general properties, each with
2 specific properties. 
Extending the
circle shown in Fig.~\ref{fig:qandy-bloch} and the Bloch sphere in the quaternionic case is
intuitive (although the geometry is far from intuitive): we obtain a
5-dimensional cube inside a 5-dimensional sphere.

\section{Multi-Qandies and Pseudo-Entanglement}\label{Sec:qandies-concepts}
We have demonstrated the principle of complementarity using the qandy picture and illustrated why the BB84 protocol is safe.
We now exploit other quantum-inspired concepts that can be demonstrated by going far beyond Jacobs' qandy model. Specifically, we shall consider here multi-qandy state, and qandies-correlations (i.e. pseudo-entanglement), closely related to quantum entanglement.

It is easy to imagine a system of multiple qandies: for example, a system of a chocolate qandy and a green qandy can be described as $\{C\}_{1} \{G\}_{2}$, where the subscript denotes the qandy's number.
The two qandies live in different ``spaces'', denoted by the separation of the two states $\{\ \}_1\{\ \}_2$. 
As we will see, we can define new types of multi-qandy systems that go beyond a group of independent qandies.
In some sense, this demonstrates another complementarity principle, as we clarify by the end of this section.

\subsection{Correlated Qandies and Pseudo-Entanglement}\label{Subsec:qandies-correlations}

Quantum entanglement is a phenomenon often called ``spooky action at a distance''. 
In its simplest form, it describes the behavior of two spatially separated systems where measurement outcomes on both systems are correlated in a weird (non-classical) way.
We can partially mimic it using two qandies.

To demonstrate this, we need to imagine a qandy-producing machine of a new type, which generates (in addition to the previously described single qandies) four types of correlated pairs of qandies. 
One type of correlated pair, denoted as $\{\phi_+\}$, has the property that if the same measurement (look or taste) is performed on both qandies, the outcomes are always random but identical.
If both are looked at, then the outcome would always be either (red, red) or (green, green), each occurring with $50\%$ probability; similarly, if both are tasted, then the outcome would always be either (chocolate, chocolate) or (vanilla, vanilla), each occurring with $50\%$ probability.
On the other hand, if one qandy is tasted while the other is looked at, each outcome is random on its own, and there are no correlations between the two outcomes. 
These correlated qandies are different than the qandies of the earlier sections in the sense that each qandy in the pair can no longer be assigned a definite taste or color, and only the correlations are now defined.

We define the other correlated qandy pairs similarly as follows:
\begin{itemize}
  \item $\{\phi_+\}$ as above; 
  \item $\{\psi_+\}$: If both candies are looked at, they present opposite colors, but whenever they are tasted, they present identical tastes;
  \item $\{\phi_-\}$: If both candies are looked at, they present identical colors, but whenever they are tasted, they present opposite tastes;
  \item $\{\psi_-\}$: If both candies are looked at, they present opposite colors, and whenever they are tasted, they present opposite tastes.
\end{itemize}

As before, in all these cases, if one qandy is tasted while the other is looked at, there are no correlations between the outcomes, and the outcomes are totally random.
These qandies possess what we call pseudo-entanglement.

We can easily mimic a variant of BB84 using such a machine. 
This, in its quantum version, is called the EPR scheme~\cite{Ekert1991quantum,bbm92}. 
Here, Alice prepares a correlated pair $\{\phi_+\}$. 
She then sends one qandy to Bob and keeps the other one. 
Each of them chooses whether to taste the qandy or look at it. 
They expect the same result if both performed the same action, and they expect random and uncorrelated results if they performed different actions. 
After repeating a sufficient number of times, they compare their actions and keep only the results from the cases in which they performed the same action. 

In all the cases where Alice and Bob performed the same action, their result is identical, random, and secure against Eve.
This ensures the safety of the protocol.
Importantly, even if Eve prepares the qandy pairs (instead of Alice), she does not know what answer Alice and Bob would get.

\subsection{Quantum Bit Commitment}\label{sec:bit-commitment}

With correlated qandies, we can show another interesting quantum cryptographic protocol; unfortunately, the protocol is actually insecure, and the simplest and most typical attack showing its insecurity can be fully described using pseudo-entangled qandies.
We first consider a goal called bit commitment, where Alice would like to first deliver a bit to Bob, and reveal the bit at a later time.
However, the two parties may not trust each other: for example, Bob may suspect that Alice could change her bit after it has been sent, and Alice may suspect that Bob could somehow know the bit before she decides to reveal it.
We divide the protocol into a commit stage and a reveal stage.

One quantum scheme (along with a proof of its insecurity!) was proposed by Bennett and Brassard in 1984~\cite{bennett1984quantum}. 
Their scheme works as follows.
In the commit stage, Alice prepares $n$ qandies with definite colors (each randomly chosen to be red or green) if she wants to commit a bit 0, or $n$ qandies with a definite taste (randomly chocolate or vanilla) if she wants to commit 1.
She keeps a record of what she has prepared (a list of red, chocolate, green, etc., for example), and delivers the $n$ qandies to Bob, without telling him how she had prepared them.

At the reveal stage, she tells Bob her choice of preparation of all $n$ qandies, which effectively uncovers the bit.
Bob, who is assumed to have a memory where he keeps the qandies\footnote{There is a similar protocol if Bob has no memory, with a slightly different analysis.}, can verify that she did not cheat (i.e., changed the bit she originally intended to send) by asking Alice what he should get for each qandy.

The safety consideration of the protocol is as follows.
Clearly, if Bob randomly looks at or tastes the qandies, the result is completely random and meaningless, as the information is fully encoded in Alice's preparation method.
Thus, the best strategy for Bob is to do nothing at the commit stage, and then he can fully verify Alice's opening at the reveal stage.
So Alice can be sure that Bob does not have any information on her bit.
On the other hand, since Alice cannot change the qandies once they're in Bob's hand, if Alice did cheat by telling Bob the opposite method, she would need to guess Bob's random results, which has a success probability of $2^{-n}$.
Therefore, cheating can be prevented with an arbitrarily high probability if Bob requires Alice to verify a longer and longer bit string.

However, the above argument does not hold if Alice is capable of preparing pseudo-entangled qandies.
Alice can cheat by doing the following --- she prepares $n$ pairs of qandies in $\{\phi_{+}\}$, sends one of each pair to Bob and keeps the other one until she wants to reveal.
At the reveal stage, she can tell Bob any bit she ``was committing'' (either color or taste), and when Bob would like to verify, she can taste or look at (according to her choice at the last minute) every qandy, and tell Bob the results to verify on the other member of each pseudo-entangled pair.
Because of the perfect correlation in both color and taste for $\{\phi_{+}\}$, Bob would always be convinced, while Alice has the complete freedom to cheat and decide what she wants to send long after the committing stage.
Therefore, the protocol is no longer safe when pseudo-entangled qandies are available to Alice; a similar insecurity argument is true in the quantum scenario and is even generalized to any possible quantum bit-commitment protocol~\cite{mayers1997unconditionally}.

This protocol and the attack demonstrate the spooky action at a distance. 
Alice controls Bob's qandies, exactly to the desired level, even though she has no access to them at the stage when she decides on how to control them.

\section{New Measurement Possibilities} \label{Sec:new-meas}
We now discuss some possibilities allowed by expanding the qandy model with new types of measurements, beyond the regular ``looking" and ``tasting" previously defined.
Namely, we introduce qandy ``Pseudo-Bell measurements", which allow us to demonstrate new QKD protocols such as the BHM96 protocol.

\subsection{A New Complementarity Principle and Pseudo-Bell Measurements} \label{sec:bell-meas}
We now define a new type of black-box machine that supports the measurement of correlations between two qandies.
Namely, a machine that \emph{cannot} look or taste, but can distinguish the four types of correlated qandy pairs -- $\{\phi_{+}\}, \{\phi_{-}\}, \{\psi_{+}\}, \{\psi_{-}\}$.

A measurement means a test of the specific property of one general property or another, namely one may look at a qandy --- measure its color, or one may taste a qandy --- measure its taste.
We called color and taste complementary to each other because when we have complete information about one general property, the uncertainty about the other general property is maximal.
For example, a qandy prepared with definite color has a completely undefined taste.
Therefore when tasted, a random taste will be observed - with equal probabilities for all the outcomes ($\{C\}$, $\{V\}$).

Applying the same principles for two qandies, we can see that the states
$\{C\}_1\{C\}_2; \{V\}_1\{V\}_2$ are complementary to 
$\{\phi_+\}; \{\phi_-\}$, in the following way:
If we prepare any of the correlated states $\{\phi_+\}; \{\phi_-\}$ and measure the color of both qandies, we may obtain $\{C\}_1\{C\}_2$ or $\{V\}_1\{V\}_2$, with equal probabilities.
And vice versa, if we prepare any of the two color-states $\{C\}_1\{C\}_2; \{V\}_1\{V\}_2$, we may define a new type of a measurement -- a correlation measurement -- that will yield the answer $\{\phi_+\}$ or $\{\phi_-\}$ with equal probabilities.
Similarly, the states $\{C\}_1\{V\}_2; \{V\}_1\{C\}_2$ are complementary to 
$\{\psi_+\}; \{\psi_-\}$. 
And one may define a measurement when $\{C\}_1\{V\}_2$ or $\{V\}_1\{C\}_2$ is prepared, that yields one of the correlation states $\{\psi_+\}$ or $\{\psi_-\}$.

Therefore, we can finally define a correlation measurement as follows.
Imagine a black box that takes a pair of qandies as input, and outputs one of the four outcomes: $\{\phi_+\}$; $\{\phi_-\}$; $\{\psi_+\}$; and $\{\psi_-\}$. 
When receiving a pair of qandies prepared in one of the four correlated states ($\{\phi_+\}$ etc.), the machine correctly identifies the type of correlation.
Therefore we have:
\begin{itemize}
	\item 
		If the two qandies are prepared with identical colors, the machine will observe $\{\phi_+\}$ or $\{\phi_-\}$, with equal probabilities.
	\item 
		If the two qandies are prepared with identical tastes, the machine will observe $\{\phi_+\}$ or $\{\psi_+\}$, with equal probabilities.
	\item 
		If the two qandies are prepared with opposite colors, the machine will observe $\{\psi_+\}$ or $\{\psi_-\}$, with equal probabilities.
	\item 
		If the two qandies are prepared with opposite tastes, the machine will observe $\{\phi_-\}$ or $\{\psi_-\}$, with equal probabilities.
\end{itemize}
Finally, if one qandy is prepared with definite color and the other with a definite taste (in other words, the two qandies are not correlated at all), all four correlation states will be observed with equal probabilities.

In quantum information, this is known as the Bell measurement.
Here, we'll call it Pseudo-Bell measurement.
Using it, one can present additional QKD protocols, as well as more amazing concepts such as qandy superdense coding.

\subsection{BHM96 QKD with Qandies} \label{sec:bhm96}
A direct application of the Pseudo-Bell measurement introduced in 
Subsection~\ref{sec:bell-meas} is the BHM96 QKD protocol \cite{bhm96} (also named reversed EPR scheme).
The protocol allows QKD between a pair of users (Alice, Bob) without transmitting qandies between them. 
This is achieved by transmitting qandies only from each user to a mutual center (Charlie) and using public classical channels.
The procedure is as follows:
\begin{enumerate}
	\item
	Preparation: Each user (Alice, Bob) generates a qandy $q \in \{R, G, C, V\}$, and remembers a classical bit associated with the qandy (same as for the BB84 protocol described in Section~\ref{BB84-qandy}) as well as the general property (color/taste).
	\item
	Transmission: Each user (Alice, Bob) sends his/her qandy to Charlie.
	\item
	Processing: Charlie applies a Pseudo-Bell measurement to both the qandies he received from Alice and Bob, and broadcasts the \emph{result} of the measurement (not the qandies themselves) to Alice and Bob.
	\item
	Post-processing: Alice and Bob compare their general properties (color/taste) via a public classical channel (which is unjammable).
	If their general properties are equal \emph{and} the result of Charlie's measurement is $\{\psi_{-}\}$ -- Bob flips his bit, and Alice and Bob now share an identical bit $b$. Else, the bits are discarded.
\end{enumerate}

The protocol can now be repeated as many times as necessary to obtain a shared secret key.
We note that out of $n$ qandies sent by Alice and Bob during the protocol, roughly 50\% are lost (their bits are discarded) due to not being of the same general property.
Out of those left, roughly 50\% are lost due to having the same specific property.
And lastly out of those left, roughly 50\% are lost due to the result not being $\{\psi_{-}\}$.
Overall, only roughly $1/8$ of the qandies contribute towards the key generation.
The efficiency can be increased by a factor of 4 if we allow using all Pseudo-Bell measurement results (and not just $\{\psi_{-}\}$).

Charlie has no way to reliably determine the value $b$ because he does not know the specific property of Alice's qandy $q$ (he only knows it is anti-correlated with Bob's qandy).
It can also be proven \cite{bhm96} the protocol is secure against a cheating Charlie -- the proof relies on the security of the EPR scheme.

\subsection{MDI-QKD with Qandies}
Suppose now Alice and Bob wish to communicate using the qandy BB84 protocol (described in Section~\ref{BB84-qandy}).
Unfortunately, Alice and Bob are both \emph{color} blind and have contracted COVID-19, impairing their sense of taste, and are thus prevented from looking at or tasting qandies on their own.
Note that Alice and Bob are still able to generate qandies using the standard qandy generating machine, as each button is labeled correctly by a letter ($\{R\}$, $\{G\}$, $\{C\}$, or $\{V\}$).

In order to communicate with Alice, Bob has purchased from Eve a special measurement device that can perform any defined qandy measurement (look/taste/Pseudo-Bell) and tell him aloud the result (lucky for Bob - he isn't deaf).
However, Alice and Bob do not trust Eve and suspect the device leaks \emph{all} measurement results to her (in addition to the usual assumption of the qandy communication being accessible to Eve).
Can we design a protocol for Alice and Bob which is measurement-device-independent (MDI), namely secure against Eve learning their secret key despite the device (and channel) being compromised?

Such a protocol is possible.
It is named MDI-QKD and relies on the BHM96 \cite{bhm96} protocol
(See \cite{xmz+20, cxc+14} for additional details and review
\footnote{In \cite{xmz+20}, a whole subsection (VI B 1) is titled ``Time-reversed EPR QKD'', showing the degree to which MDI-QKD (hence also MDI quantum cryptography in general) relies on the reversed-EPR protocol.}).
Having defined the BHM96 protocol in Section~\ref{sec:bhm96}, it is not difficult to see we can merge Bob's and Charlie's roles in the BHM96 protocol to obtain a protocol with the required properties.
Instead of Charlie being the untrusted center, Bob uses his untrusted measurement device to fulfill Charlie's role. 
Alice will send her qandy to Bob instead of Charlie, and Bob will Pseudo-Bell measure his and Alice's qandies and report the result to Alice.
The rest of the protocol is unchanged.
The security of the protocol against Eve's measurement device follows directly from the security of the BHM96 protocol.
The security against the channel being compromised may be achieved by following techniques similar to the BB84 protocol (comparing TEST bits, Error-Correction and Privacy Amplification).

\section{Qandy Gates}\label{Sec:qandy-gates}

Quantum gates are the basic units describing \emph{transformations} in quantum information.
In the qandy world, a gate can be thought of as a black box with inputs and outputs, where some qandies are sent in and some (potentially different) qandies are produced based on what the inputs are.
The full effect of a gate is captured by a table relating the inputs and outputs.

It's important to note that what happens inside the black box is not the action of a hidden observer who tries to observe the input qandy first and then produces some new qandies using the machine in Figure~\ref{candy-fig}.
This is because if the observer does not know what the input qandy is, half of the time he/she would be observing using the wrong method and revealing the wrong property.
Instead, a gate should be thought of as applying the same operation regardless of the input, and different qandies react differently to the same gate.

\subsection{Qandy Gates on Single Qandies} \label{sec:single-gates}
We now define several simple gates on single qandy inputs producing single qandy outputs, inspired by their direct counterparts in quantum computing.
To start, a color-switching gate is a gate that on an input of a red or green qandy outputs a qandy with the opposite color, but would leave a chocolate or vanilla qandy unaltered.
This represents a Pauli X gate in quantum computing.
Similarly, we define a taste-switching gate, corresponding to a Pauli Z gate.
These gates can be succinctly described by a set of input-output rules, as seen in Table~\ref{tab:qandy-gates}.

\begin{table}[!ht]
	\centering
	\begin{tabular}{ |c|c|c| } 
		\hline
		Input&$X$&$Z$\\
		\hline
		$\{R\}$&$\{G\}$&$\{R\}$\\
		$\{G\}$&$\{R\}$&$\{G\}$\\
		$\{C\}$&$\{C\}$&$\{V\}$\\
		$\{V\}$&$\{V\}$&$\{C\}$\\
		\hline
	\end{tabular}
	\caption{Input-output rules for a qandy X (color-switching) and qandy Z (taste-switching) gates}
	\label{tab:qandy-gates}
\end{table}

\subsection{Qandy Gates on Correlated Qandy Pairs} \label{sec:gates-on-bell}
The definitions of the gates in Table~\ref{tab:qandy-gates} assume the input qandies are single, uncorrelated qandies with a defined state.
However, for an input qandy which is a member of a correlated qandy pair - namely a qandy from a pair $\{\phi_+\}$; $\{\phi_-\}$; $\{\psi_+\}$; or $\{\psi_-\}$ - the action of the gates we described is generally undefined.

However, for some gates - such as the color-switching qandy X gate and the
taste-switching qandy Z gate, it makes sense to define their action on a member
of a correlated qandy pair (and as we'll see such definitions will prove useful
when we describe the qandy superdense coding protocol in 
Subsection~\ref{sec:superdense}).

For example, consider a correlated qandy pair in the state $\{\phi_+\}$, namely a pair of qandies with identical colors when looked at and identical tastes when tasted.
If we generalize the action of the color-switching gate defined in Table~\ref{tab:qandy-gates}, and assume its application on a member of a pair does not break the correlation - it is natural to define the state of the resulting pair of qandies to have opposite colors when looked at and identical tastes when tasted.
This is because the color-switching gate only switches colors and not tastes, hence we can compare the measurement result before and after applying the gate (for example on the left qandy) as described in Table~\ref{tab:gates-bell-res}.

\begin{table}[!ht]
	\centering
	\begin{tabular}{ |c|c|c| } 
		\hline
		Measurement&Result before gate&Result after gate\\
		\hline
		Look at both&$\{R\}$,$\{R\}$&$\{G\}$,$\{R\}$\\
		Look at both&$\{G\}$,$\{G\}$&$\{R\}$,$\{G\}$\\
		Taste both&$\{C\}$,$\{C\}$&$\{C\}$,$\{C\}$\\
		Taste both&$\{V\}$,$\{V\}$&$\{V\}$,$\{V\}$\\
		\hline
	\end{tabular}
	\caption{Possible measurement results before and after applying a qandy X (color-switching) gate on the left member of a $\{\phi_+\}$ correlated pair}
	\label{tab:gates-bell-res}
\end{table}

We can see according to Table~\ref{tab:gates-bell-res} that the state after applying the color-switching gate is a $\{\psi_+\}$ correlated pair.
In a similar fashion, we can generalize the action of the color-switching X gate on all correlated qandy pair states, as well as the action of the taste-switching Z gate, to arrive at the description in Table~\ref{tab:gates-on-bell}.

\begin{table}[!ht]
	\centering
	\begin{tabular}{ |c|c|c| } 
		\hline
		Input&$X$&$Z$\\
		\hline
		$\{\phi_+\}$&$\{\psi_+\}$&$\{\phi_-\}$\\
		$\{\phi_-\}$&$\{\psi_-\}$&$\{\phi_+\}$\\
		$\{\psi_+\}$&$\{\phi_+\}$&$\{\psi_-\}$\\
		$\{\psi_-\}$&$\{\phi_-\}$&$\{\psi_+\}$\\
		\hline
	\end{tabular}
	\caption{Input-Output rules for qandy X and qandy Z gates acting on any (single) member of a correlated qandy pair}
	\label{tab:gates-on-bell}
\end{table}

\subsection{Qandy Superdense Coding} \label{sec:superdense}
We now give an example for a communication protocol which uses qandy gates (as defined in Section~\ref{Sec:qandy-gates}) together with Pseudo-Bell measurements (as defined in Section~\ref{sec:bell-meas}).
The protocol allows a message of 2 classical bits to be encoded and transmitted via a single qandy.
The procedure is as follows:
\begin{enumerate}
	\item 
	Preparation: Charlie generates the $\{\phi_{+}\}$ correlated qandy state, sends one of the qandies to Alice and the other to Bob.
	\item 
	Encoding: Alice chooses 2 classical bits $b_1, b_0 \in \{0, 1\}$ she would like to send to Bob, and applies the single qandy gates on her qandy described by Table~\ref{tab:superdense-enc}.
	\begin{table}[!ht]
		\centering
		\begin{tabular}{ |c|c|c| } 
			\hline
			$b_1$&$b_0$&Gates applied\\
			\hline
			0&0&$I$\\
			0&1&$X$\\
			1&0&$Z$\\
			1&1&$X$ and $Z$\\
			\hline
		\end{tabular}
		\caption{Qandy superdense coding encoding stage: actions by Alice based on the bits she wants to transmit}
		\label{tab:superdense-enc}
	\end{table}
	Note that in case of $(b_1, b_0) = (1,1)$ the order of the application of $X,Z$ does not matter.
	\item 
	Transmission: Alice sends her \emph{single} qandy to Bob
	\item 
	Decoding: Bob measures both the qandies using a Pseudo-Bell measurement, and interprets his results as described by Table~\ref{tab:superdense-dec}.
	\begin{table}[!ht]
		\centering
		\begin{tabular}{ |c|c|c| } 
			\hline
			Measurement result&$b_1$&$b_0$\\
			\hline
			$\{\phi_{+}\}$&0&0\\
			$\{\psi_{+}\}$&0&1\\
			$\{\phi_{-}\}$&1&0\\
			$\{\psi_{-}\}$&1&1\\
			\hline
		\end{tabular}
		\caption{Qandy superdense coding decoding stage: bits decoded by Bob according to his measurement result}
		\label{tab:superdense-dec}
	\end{table}
\end{enumerate}
The correctness of the protocol follows from the definition of qandy $X$ (color-switching) and $Z$ (taste-switching) gates and their actions on the correlated states (as defined in Section~\ref{sec:gates-on-bell}).
For example, when Alice applies only a $X$ gate on her qandy in the $\{\phi_{+}\}$ correlated pair, the state is changed to $\{\psi_{+}\}$, which Bob will then measure and correctly conclude $(b_1, b_0) = (0, 1)$.
Similarly, when Alice applies both $X$ and $Z$, starting from $\{\phi_{+}\}$ she will arrive at the state $\{\psi_{-}\}$ (and the order of application does not matter, as can be clearly seen from Table~\ref{tab:gates-on-bell}).
Bob will then measure $\{\psi_{-}\}$ and conclude $(b_1, b_0) = (1, 1)$.
The other two cases are analogous.

\subsection{Multi-Qandy Gates and Qandy CNOT Gate} \label{sec:qandy-cnot}
Similar to the single qandy gates defined in Section~\ref{sec:single-gates}, and in direct analogy to the quantum mechanical definitions, we can define multi-qandy gates.
A multi-qandy gate takes $n \geq 2$ qandies as input and produces $n$ qandies as an output.
(We define the number of output qandies to be equal to the number of input qandies because the multi-qandy gates must be reversible - in order to be consistent with quantum mechanics).

A particularly useful gate often used by an adversary in many QKD protocols is the CNOT gate.
We define a qandy CNOT (taste) gate, which intuitively acts as follows on a pair of uncorrelated qandies:
If the first qandy (dubbed ``control") is $\{C\}$, do nothing.
Else, apply the taste-switching ($Z$) gate on the second qandy (dubbed ``target").

The partial action of the qandy CNOT (taste) gate, when the control qandy has a specific color, can be summarized in Table~\ref{tab:qandy-cnot}.
\begin{table}[!ht]
	\centering
	\begin{tabular}{ |c|c| } 
		\hline
		Input&CNOT\\
		\hline
		$\{C,R\}$&$\{C,R\}$\\
		$\{C,G\}$&$\{C,G\}$\\
		$\{C,C\}$&$\{C,C\}$\\
		$\{C,V\}$&$\{C,V\}$\\
		$\{V,R\}$&$\{V,R\}$\\
		$\{V,G\}$&$\{V,G\}$\\
		$\{V,C\}$&$\{V,V\}$\\
		$\{V,V\}$&$\{V,C\}$\\
		\hline
	\end{tabular}
	\caption{Partial input-output rules for a qandy CNOT (taste) gate, when the control qandy is a specific-taste qandy}
	\label{tab:qandy-cnot}
\end{table}

We note that in order to fully define the qandy CNOT (taste) gate, we also need to specify its action in the following cases:
\begin{enumerate}
	\item 
		A pair where the control qandy has a specific color
	\item 
		The two qandies form a correlated pair
\end{enumerate}
Such definitions may be taken straightforwardly and analogously to the case of the quantum CNOT gate, similar to our definition of the single qandy gates on correlated pairs in Section~\ref{sec:gates-on-bell}.
Thus, we avoid giving a full definition of the qandy CNOT (taste) gate, and henceforth simply assume the partial definition given in Table~\ref{tab:qandy-cnot} can be extended in a reversible manner.
Extending the definition as described we obtain that the qandy CNOT gate is its own inverse.

\section{Semi-Quantum Cryptography with Qandies}\label{Sec:SQKD}

We now discuss in further detail the semi-quantum key distribution (SQKD) protocols, which we briefly mentioned in Section~\ref{sec:trivial-exten}.
We start by mentioning that following the first paper on SQKD \cite{bkm07}, other cryptographic protocols were shown with a classical party \cite{ZQLWL09, BM11, LC2008, SDL2013, YYLH2014, Kra15}.
In this section, we focus solely on SQKD as discussed in \cite{bkm07, bgkm09}, and explain these ideas using qandies instead of qubits.
Namely, we discuss QKD with the ``color-blind" Bob constraint - Bob may only taste a qandy, prepare a qandy with a specific taste or reflect a qandy back to Alice, undisturbed.

At first, we might be tempted to define a ``Mock SQKD Protocol".
The protocol is defined as follows:
\begin{enumerate}
	\item 
		Alice sends a qandy $\{R, G, C, V\}$ to Bob, as in the BB84 protocol.
	\item 
		Bob randomly decides to either taste the qandy or reflect it back to Alice.
	\item 
		In case Bob decided to reflect the qandy, Alice measures it (looks/tastes) according to how she prepared the qandy.
	\item 
		Steps 1-3 are repeated for as many iterations as needed. After the final iteration, Alice will publish which qandies she prepared with a specific taste.
\end{enumerate}	
The bits obtained from the qandies which Alice sent with specific taste and Bob decided to taste form the ``Sifted Key", which Alice and Bob can then use to obtain a secret key as in the BB84 protocol.
For the qandies reflected by Bob, Alice expects to measure the exact specific property she originally prepared.
Hence these qandies are used by Alice to check for the presence of eavesdropping activity on the Qandy channel between her and Bob.
However, motivated by the quantum protocol, the ``Mock SQKD protocol" is insecure, if we follow operations that Eve is allowed to perform in the quantum protocol: an adversarial Eve can use the qandy CNOT gate (defined in Section~\ref{sec:qandy-cnot}) to obtain full information about Alice and Bob's shared key without being detected, which we will discuss in Section~\ref{sec:mock-cnot-attack}.

Due to the weakness of the ``Mock Protocol", noticed already in~\cite{bkm07}, 
alternative SQKD protocols were suggested in \cite{bkm07, bgkm09}.
In the first SQKD protocol, the ``Taste-Resend" protocol \cite{bkm07}, we require that in addition to the minimal set of operations (of tasting a qandy and reflecting it), Bob is able to generate a new qandy with a specific taste $\{C, V\}$.
The ``Taste-Resend" protocol is identical to the ``Mock Protocol", with the following difference: in step 2, instead of simply tasting the qandy - Bob will taste the qandy, create a qandy with the same taste he measured and send it back to Alice.
After the final iteration, Alice publishes which qandies she prepared with a specific taste, and Bob publishes which qandies he tasted and resent and which qandies he just tasted.

An alternative SQKD protocol was given in \cite{bgkm09}, in which the requirement on Bob to be able to generate specific-taste qandies is relaxed.
The significance of this protocol is in the fact that in real world QKD systems, it is sometimes difficult to ``Measure-Resend" a received qubit.
The protocol is almost identical to the first SQKD protocol presented above, with the following key differences:
\begin{enumerate}
	\item 
		Alice first sends all of her qandies to Bob, and only then Bob acts on them
	\item 
		Bob does not resend the qandies he decides to measure and instead randomizes the order of the qandies he decides to reflect (Bob needs to be able to ``delay" reflected qandies)
	\item 
		Alice collects all qandies reflected by Bob in a ``Qandy Memory"
	\item 
		Instead of publishing which qandies he tasted and resent and which he just tasted, Bob publishes which qandies he reflected (and in which order) and which he just tasted.
\end{enumerate}
Both protocols were proven to be completely robust, namely that any information on the shared key acquired by an adversarial Eve implies a non-zero probability that she will be detected by Alice and Bob \cite{klm20}.


\subsection{The CNOT Attack on the ``Mock SQKD" Protocol} \label{sec:mock-cnot-attack}

We have seen the ``Mock SQKD" protocol and claimed it is insecure against a particular eavesdropping attack by an adversary.
With the definition of qandy CNOT (taste) gate in place 
(Subsection ~\ref{sec:qandy-cnot}), 
we may now present the CNOT attack on the ``Mock SQKD" protocol.

Assuming Eve possesses a qandy CNOT (taste) gate as described, she can hold an ancilla $\{C\}$ qandy for each qandy sent from Alice to Bob.
When Alice sends a qandy to Bob, Eve will intercept the qandy and apply the qandy CNOT (taste) gate on the intercepted qandy (as a ``control") and on her ancilla qandy (as a ``target").
She then stores her ancilla qandy, forwards the intercepted qandy to Bob, and attempts to intercept his response.
There are now two scenarios, according to Bob's decision:
\begin{enumerate}
	\item If Bob decides to reflect the qandy, Eve will simply intercept the reflected qandy, apply the qandy CNOT (taste) gate again on Bob's qandy as ``control" and her stored ancilla as ``target" and forward the intercepted qandy to Alice.
	Due to the qandy CNOT gate being its own inverse as mentioned in \ref{sec:qandy-cnot}, the qandy will return back to its original state, with Alice unable to detect Eve's presence when she measures the qandy.
	\item Otherwise, namely if Bob decides to taste his qandy, Eve may taste her ancilla qandy.
	There are now again two scenarios:
	\begin{enumerate}
		\item If Alice has prepared the qandy with a specific taste, Alice and Bob will use the bit value for their ``Sifted Key".
		All three parties share the same bit value and Eve has full information about the shared key.
		\item Otherwise, namely Alice has prepared the qandy with a specific color, Bob's and Eve's taste measurements will be random (but correlated\footnote{A comment for the experts: Straightforwardly extending the definition of the qandy CNOT gate, we obtain that Bob and Eve share a correlated pair - $\phi_{+}$ or $\psi_{+}$, depending on the qandy Alice prepared.}).
		Unfortunately, this doesn't help to detect Eve's presence, since even without Eve's intervention Bob would measure a random taste (with the same probabilities) in this case.
	\end{enumerate}
\end{enumerate}
This attack clarifies why the original ``Mock SQKD" protocol does not work and gives intuition on why the additional steps of ``Taste-Resend" or randomization are required.

\section{Non-Local Boxes With Qandies}\label{Sec:NLB}
In this section, we demonstrate how the qandy model can be extended to include and explain certain phenomena beyond quantum theory -- Non-Local Boxes (NLBs).
Thus the qandy model may be seen not only as a ``slim'' version of quantum theory, but also as a general tool for describing other non-trivial models.

\subsection{The CHSH Game and Non-Local Boxes}
The concept of NLBs was first proposed by Popescu and Rohrlich \cite{pr94}, and can be described as a solution to the CHSH game \cite{chsh69, bm06}.

The CHSH game (as described in \cite{bm06}) is a game played between Alice and Bob on one side, and Charlie on the other.
Charlie generates 2 random bits $x_A, x_B \in \{0,1\}$ and sends one to Alice and the other to Bob.
Alice and Bob's goal is to produce two output bits $y_A, y_B \in \{0, 1\}$ (one from Alice and one from Bob) such that the following conditions holds:
\begin{equation} \label{eq:csch_game}
	x_A \text{ AND } x_B = y_A \text{ XOR } y_B.
\end{equation}
Before the game starts, Alice and Bob are allowed to meet and devise a strategy, but afterwards (and before Charlie sends them the bits) they are separated and can no longer communicate.
Can Alice and Bob come up with a strategy to win the game?

Classically, it turns out that no matter what strategy Alice and Bob come up with, they can win only 75\% of the time.
This is evident by describing all possible strategies via pairs of deterministic functions $f_A, f_B : \{0,1\} \rightarrow \{0,1\}$ which Alice and Bob use to select their answer, based on the input bit.
If Charlie sends uniformly random bits, with each of the deterministic strategies at least one of the 4 possible inputs $x_A, x_B$ will cause a loss of the game.
Any probabilistic strategy can be seen as a probability distribution of the appropriate deterministic strategies, and thus can do no better than the best deterministic strategy.

If Alice and Bob are allowed to use Quantum Mechanics, it can be shown they can devise a strategy to win with probability $\cos^2(\pi/8) \approx 0.85$.
They do so by preparing a Bell pair $\ket{\phi_+} = (\ket{00}+\ket{11})/\sqrt{2}$ when they meet, and then make appropriate measurements on their qubits according to the input bits (and report the measurement result as their answer).
This solution was proposed by Clauser, Horne, Shimony and Holt \cite{chsh69}, after whom the game is named.
Later, this strategy was shown to be optimal in the quantum case, via the Cirel'son Inequality \cite{cir80}.

In \cite{pr94}, Popescu and Rohrlich have explored what is possible by taking quantum non-locality as an axiom, while preserving relativistic causality (namely, Alice cannot communicate any information to Bob just by measuring, and vice-versa).
Somewhat surprisingly, they found that even under such constraints, it is possible to define a device shared between Alice and Bob, with each of them having access to a single bit input and a single bit output, such that eq.~\eqref{eq:csch_game} is satisfied.
This stands in stark contrast to what is allowed by quantum mechanics via the Cirel'son Inequality, and suggests a ``Beyond Quantum'' theory.

Following \cite{pr94}, the concept of NLBs was further researched and analyzed, with some milestone results.
In \cite{cgmp05}, it was shown that NLBs as suggested in \cite{pr94} can be used to simulate any projective measurement on the singlet Bell pair $\ket{\psi_{-}}=(\ket{01}-\ket{10})/\sqrt{2}$.
Then, in \cite{bm06}, it was shown there exist quantum correlations which cannot be simulated efficiently by NLBs, and vice-versa.
Lastly, in \cite{van13}, it was shown that NLBs allow any distributed computation to be made with single bit communication;
since such a scenario is unlikely, this result suggests a limitation on the existence of NLBs in nature.

\subsection{Non-Local Qandies}
In the qandy world, we can easily envision the original 
Popescu-Rohrlich NLB, by defining a machine that produces 
a pair of correlated qandies $\{\chi\}$ as follows: 
if both qandies are tasted, or one of them is looked at and 
the other is tasted - the results are identical 
(for some appropriate notion of identity between color and taste), 
whereas if both qandies are looked at they appear with opposite colors. 
See Table~\ref{tab:nlb-vs-phi} for a comparison between the 
pseudo-Bell pair $\{\psi_+\}$ (defined in 
Section~\ref{Subsec:qandies-correlations})
and the non-local qandy pair $\{\chi\}$).

\begin{table}[!ht]
	\centering
	\begin{tabular}{ |c|c|c| } 
		\hline
		Action&$\{\psi_+\}$&$\{\chi\}$\\
		\hline
		Taste, Taste&Identical&Identical\\
		Taste, Look&Random&Identical\\
		Look, Taste&Random&Identical\\
		Look, Look&Opposite&Opposite\\
		\hline
	\end{tabular}
	\caption{The results of various measurements on the pseudo-Bell pair $\{\psi_+\}$ vs. the non-local qandy pair $\{\chi\}$.
		Note the difference between the pairs when the measurements are not identical (tasting one and looking at the other).}
	\label{tab:nlb-vs-phi}
\end{table}

Of course, Alice and Bob can use such a machine to win the CHSH game with certainty:
they create a non-local qandy pair using the machine, each of them puts a qandy in their pocket and they begin the game.
When Charlie sends the bits $x_A$ and $x_B$, Alice and Bob will either taste their qandy if the bit is $0$, or look at their qandy if the bit $1$, and report their measurement result as the output bits $y_A$ and $y_B$, respectively.
By definition of the non-local qandy pair $\{\chi\}$, their outcomes will satisfy eq.~\eqref{eq:csch_game}.

Following this, one can define other types 
of NLBs using qandies in a similar fashion:
for example, any permutation on the qandy NLB suggested above, i.e. a pair of correlated qandies that always produce identical results, unless both are tasted, or unless one is tasted and the other is looked at (and vice-versa), in which case they produce opposite results.
Equally, an NLB can be defined that always produces opposite measurement results, unless both qandies are looked at, presenting identical colors.
Generalizing this, we may have a total of $2^4 = 16$ such qandy NLBs - by choosing an outcome - identical or opposite - to each of the 4 possible measurements (\{Look, Look\}, \{Look, Taste\}, \{Taste, Look\}, \{Taste, Taste\}).

Comparing the two columns of Table~\ref{tab:nlb-vs-phi}, we immediately
see that one can define a special type of NLBs in which random
results appear only once in the four possible locations,
rather than not appearing at all as in a regular NLB or appearing twice
as in quantum fully-entangled states.
Curiously, a ``super candy machine'' with just two buttons could, per demand, generate either an NLB qandy pair or a pseudo-entangled qandy pair.

As a final remark, we point out that other types of NLBs may be defined beyond the Popescu-Rohrlich type (e.g., a multi-party NLB described in \cite{bm06}), and we leave the exploration of such options to subsequent works.

\section{Conclusion} \label{Sec:Discussion}
In this work, we have introduced and largely extended the quantum 
candy model proposed by Jacobs and used it to demonstrate some 
of the most important concepts in quantum information science 
and quantum cryptography in an approachable manner.
The preliminary version of this paper (by J.L.~and T.M.), 
that appeared in \cite{lm20}, already presented pairs of correlated qandies
and a qandies' machine that can generate such correlations.
This work extends the qandies model much beyond the preliminary version.
The main
extensions are: to include
correlated-qandies measuring device, to present protocols (and attacks) 
relying on that, and to discuss possibilities
that go far beyond quantum theory in   
Sections~\ref{Sec:nontrivial-exten} and~\ref{Sec:NLB}. 

Our work is meant to be a useful tool for introducing quantum science 
to the general public, while also being intriguing to the experts.
We emphasize that the qandy model does not always need to follow quantum theory and can be further extended in many ways.
We leave a deeper analysis of these extensions, 
as well as a rigorous connection to quantum theory
(which mainly contains quantum superposition), to future works on the topic.

\section*{Acknowledgments}J.L. and T.M. thank the Schwartz/Reisman Foundation. J.L. is supported by NSERC Canada. T.M. and R.S. were also partially supported by Israeli MOD.